# Structural disorder in $Li_x(C_5H_5N)_yFe_{2-z}Se_2$ and $Cs_xFe_{2-z}Se_2$ superconductors studied by Mössbauer spectroscopy

K. Komędera[1], A. K. Jasek[1], A. Błachowski[1], K. Ruebenbauer[1], J. Żukrowski[2], A. Krztoń-Maziopa[3], and K. Conder[4]

[1]Mössbauer Spectroscopy Laboratory, Pedagogical University
ul. Podchorążych 2, PL-30-084 Kraków, Poland

[2]AGH University of Science and Technology,
Academic Center for Materials and Nanotechnology
Av. A. Mickiewicza 30, PL-30-059 Kraków, Poland

[3]Warsaw University of Technology, Faculty of Chemistry
ul. Noakowskiego 3, PL-00-664 Warsaw, Poland

[4]Laboratory for Developments and Methods, Paul Scherrer Institut
CH-5232 Villigen PSI, Switzerland

[*]Corresponding author: sfrueben@cyf-kr.edu.pl

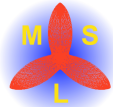

## Abstract

Two iron-chalcogenide superconductors $Li_x(C_5H_5N)_yFe_{2-z}Se_2$ and $Cs_xFe_{2-z}Se_2$ in the as-prepared and annealed state have been investigated by means of the Mössbauer spectroscopy versus temperature. Multi-component spectra are obtained. One can see a non-magnetic component due to iron located in the unperturbed Fe-Se sheets responsible for superconductivity. Remaining components are magnetically ordered even at room temperature. There is some magnetically ordered iron in Fe-Se sheets perturbed by presence of the iron vacancies. Additionally, one can see iron dispersed between sheets in the form of magnetically ordered high spin trivalent ions, some clusters of above ions, and in the case of pyridine intercalated compound in the form of α-Fe precipitates. Pyridine intercalated sample shows traces of superconductivity in the as-prepared state, while cesium intercalated sample in the as-prepared state does not show any superconductivity. Superconductors with transition temperatures being 40 K and 25 K, respectively, are obtained upon annealing. Annealing leads to removal/ordering of the iron vacancies within Fe-Se sheets, while clusters of α-Fe grow in the pyridine intercalated sample.

## 1. Introduction

Iron-based superconductors are characterized by the layered structure having some similarity to the superconducting cuprates. The layer is composed of the iron atoms and some pnictogen or chalcogen atoms. The layer is more corrugated than corresponding layer of cuprates. One can substitute some iron or pnictogen/chalcogen atoms by similar atoms [1]. For some compounds there are many iron vacancies within the layer and the vacancies order at lower temperatures [2]. Layers are stacked either directly or in some cases with inversion forming generally tetragonal structure undergoing some orthorhombic distortion in some cases and at low temperature. A separation between layers is controlled by introduction of some "spacers" except for the simplest structures. Many atoms or molecules could serve as spacers including rare earth ions with localized magnetic moments. The spacers could be diluted replacing them by various atoms of similar chemistry. For majority of structures (except simplest) the parent compound is a metallic highly anisotropic system developing itinerant magnetic order of the spin density wave (SDW) at low temperature usually accompanied by the orthorhombic distortion above mentioned [3, 4]. Various kinds of SDW are seen. Namely, one observes longitudinal SDW for '122' pnictogen family [5], while in the case of selenides/tellurides one sees in some cases a transversal SDW with a tendency to form circular or elliptical spin order [6]. All these orders are of the anti-ferromagnetic character and in some cases there are several SDW types of order forming stripes. Rare earth atoms order anti-ferromagnetically at much lower temperatures (especially those having large orbital contribution to the magnetic moment) and the coupling between two magnetic systems is weak, albeit present [7]. In order to obtain superconductivity one has to destroy SDW order, but not necessarily the rare earth magnetic order [7]. This goal could be achieved by doping of any element or pressure. Hence, one can change the Fermi surface either by electron or hole doping. The isovalent destruction of SDW is possible, too, by variation of the unit cell parameters. In the case of selenides there are some anti-ferromagnetic Mott-type insulators very close to the metallic systems developing SDW and prone to the superconducting transitions [4]. Some chalcogen-based compounds contain interstitial iron between sheets with the localized and large magnetic moment [7]. This iron has to be at least partly removed in order to achieve superconductivity [8]. Some systems develop complex charge density waves (CDW) in the superconducting state composed particularly of electrons having non-zero angular momentum [9]. An orthorhombic distortion is generally absent for systems free of SDW. Hence, one can conclude that magneto-elastic effects are important for this class of compounds.

A density of the superfluid is generally low even for compounds having high transition temperatures to the superconducting state. One can observe a large difference between the first and second critical field for these second kind superconductors. Hence, it is believed that the superfluid tends to concentrate in the iron layers with exceptionally high two-dimensional density [10]. Therefore the strength of coupling between layers is critical for observed transition temperatures. On one hand, the weak coupling increases two-dimensional superfluid density leading to enhancement of the transition temperature. On the other hand, one has to keep some electrical coupling between layers to achieve bulk superconductivity and stability of the structure. Generally, one wants to keep rather large distance between layers. Hence, various atoms or molecules were used to intercalate iron-selenium sheets with the ordered iron vacancies for the '122' like family. One can try large alkaline metal atoms like e.g. cesium [11] or try some other molecules like e.g. $NH_3$ [12-14]. Pyridine ($C_5H_5N$) seems a good choice provided it is doped by some electron donors like lithium or other alkali metals. Lithium is a small atom easily fitting between large aspherical organic molecules. Anhydrous solution of lithium in pyridine could be intercalated from the solution to the tetragonal compound bearing iron-selenium sheets with ordered iron vacancies. One has to anneal the sample subsequently to intercalation to order aspherical spacing molecules and increase order disturbed by intercalation. Therefore one can obtain superconducting material with high transition temperature provided vacancies in the iron sheet are ordered to prevent breaking of the Cooper pairs [15]. Somewhat similar effect is obtained upon intercalation of Cs, however a transition temperature is much lower. The annealing is essential as well, as the original intercalation by Cs seems to leave significant disorder between iron bearing sheets [11]. Recently it has been suggested that for Cs intercalated compound one can expect in addition to the ordered iron vacancies some Cs vacancies ordered, too. In the case of iron vacancies one obtains some complex anti-ferromagnetic order, while the superconductivity is obtained for the lack of iron vacancies and presence of the ordered Cs vacancies. The two phases coexist, albeit they are spatially separated [16].

We have performed $^{57}$Fe iron Mössbauer spectroscopy versus temperature on polycrystalline samples of $Li_x(C_5H_5N)_yFe_{2-z}Se_2$ and powdered single crystalline $Cs_xFe_{2-z}Se_2$ compounds. Above materials were investigated in the as-prepared form and upon being annealed.

## 2. Experimental

Preparation of the samples $Li_x(C_5H_5N)_yFe_{2-z}Se_2$ and $Cs_xFe_{2-z}Se_2$ together with the annealing process is described in Refs [15] and [11], respectively. Alkali metal intercalated samples were obtained via room temperature intercalation into the FeSe matrix in pyridine solutions of the corresponding alkali metals. Iron selenide (FeSe) was synthesized from high purity (at least 99.99 %, AlfaAesar) powders of iron and selenium according to the procedure described elsewhere (for details see Ref. [15]). For the intercalation process an appropriate amount of a powdered FeSe precursor was placed into a container filled with 0.2 M solution of pure alkali metal dissolved in anhydrous pyridine. The amount of FeSe taken for intercalation was calculated for the molar ratio 1:2 of alkali metal and precursor, respectively. The reaction was carried out at 40 °C until the discoloration of the alkali metal solution. After synthesis the intercalated material was separated from the solution, washed repeatedly with fresh pyridine and dried up to a constant mass in inert atmosphere. All the work was performed in a He-filled glove box to protect the powder from oxidation. Single crystals of cesium intercalated iron selenides were grown from the melt using the Bridgman method. Ceramic rods of the iron selenide starting material were prepared by the solid state reaction technique (for details see Ref. [15]). The nominal stoichiometry of the starting material that is $FeSe_{0.98}$ was chosen based on the view of our previous studies (see [15]) which demonstrated that for this particular Fe/Se ratio the content of secondary phases is the smallest. For the single-crystal synthesis a piece of the ceramic rod of $FeSe_{0.98}$ was sealed in a double-wall evacuated silica ampoule with the pure alkali of at least 99.9 % purity (Chempur). The ampoules were annealed at 1030 °C for 2 h to assure homogenization. The melt was cooled down to 750 °C at the rate of 6 °C/h. Subsequently the sample was cooled down to room temperature at the temperature rate amounting to 200 °C/h. Crystal-rods having diameter of 7 mm (diameter of the quartz ampoule) were obtained. They could be easily cleaved into plates with flat shiny surfaces [11]. For annealing process samples (single crystals or pellets of the polycrystalline material) were sealed with a flame in evacuated quartz tubes and annealed over 60 h at 215 °C for $Li_x(C_5H_5N)_yFe_{2-z}Se_2$ and at 210 °C for $Cs_xFe_{2-z}Se_2$.

The phase purity of the prepared materials was determined by powder X-ray diffraction (XRD) using a D8 Advance Bruker AXS diffractometer with Cu Kα radiation using low background airtight sample holder to avoid sample degradation during measurements. Powder

X-ray diffraction patterns for the as-prepared and annealed Li$_x$(C$_5$H$_5$N)$_y$Fe$_{2-z}$Se$_2$ sample are shown in Figure 1. Crystal structure of the main phase of the as-prepared material is well described with tetragonal structure fitted to *P4/mmm* crystal metric with *a* = 3.5911(8) and *c* = 16.3979(7) Å. The as-prepared sample also contains some FeSe as an impurity phase. Upon annealing the unit cell expands significantly and the (002) reflection moves from $2\Theta$ = 10.8° to 7.6°, but the crystal structure is preserved and can be well fitted also to *P4/mmm* with enhanced lattice parameters: *a* = 9.9047(3) and *c* = 23.1120(6) Å.

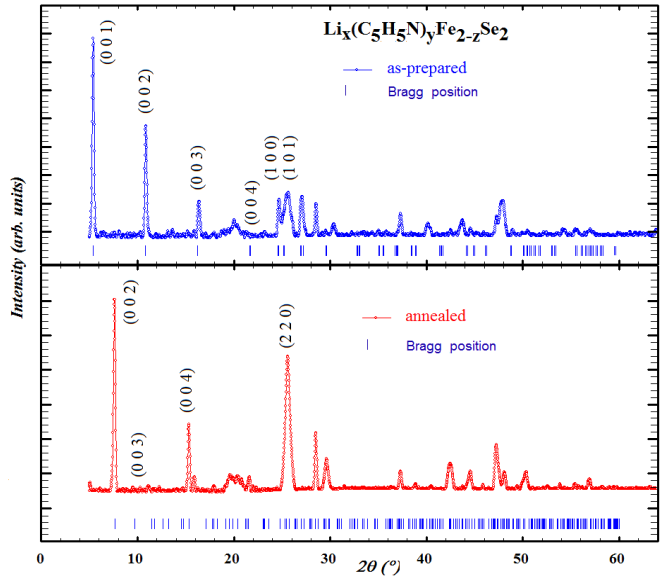

**Figure 1** Powder X-ray pattern of the as-prepared and annealed sample of Li$_x$(C$_5$H$_5$N)$_y$Fe$_{2-z}$Se$_2$. A Cu Kα radiation was used to obtain X-ray pattern. The first diffraction peaks of the main phase are labeled.

Cs$_x$Fe$_{2-z}$Se$_2$ crystal has been prepared directly from elements using the conventional Bridgman technique. After synthesis the material was examined with X-ray diffraction and micro-XRF fluorescence to check the phase purity and elemental composition. The total (average) composition of the material after synthesis was Cs$_{0.77}$Fe$_{1.59}$Se$_2$ and remained almost unchanged after prolonged annealing. The annealing temperature for Cs-122 samples was chosen on the basis of differential calorimetry measurement (shown in Figure 2) indicating the available temperature window for sample annealing, which must be realized above the phase separation temperature ($T_p$) and below the structural rearrangement temperature ($T_s$). X-ray diffraction patterns for the Cs$_x$Fe$_{2-z}$Se$_2$ sample in the as-prepared and annealed state are shown in Figure 3. An inset of Figure 3 shows the evolution of the (002) reflection upon prolonged annealing of the sample in evacuated quartz ampoule. The post-annealing of the as-prepared Cs-crystal induces the micro-structural changes leading to the evolution towards the superconducting state.

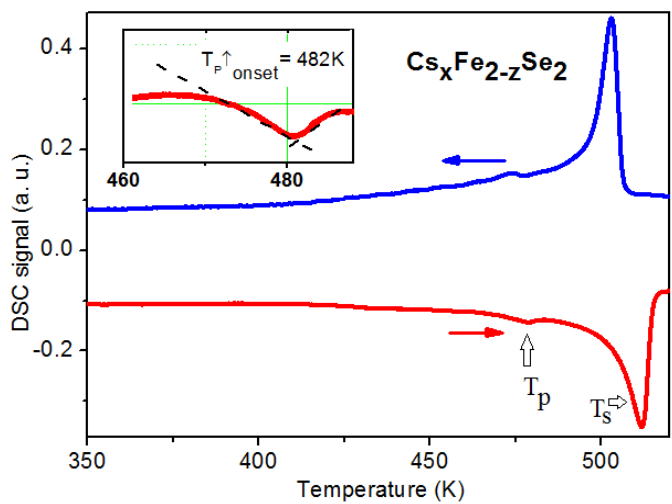

**Figure 2** DSC signal versus temperature recorded for Cs$_x$Fe$_{2-z}$Se$_2$ sample on heating with 20 K/min (red line) and cooling (blue). $T_p$ indicates the peak corresponding to the nanoscale phase separation temperature and $T_s$ stands for the vacancy order-disorder temperature.

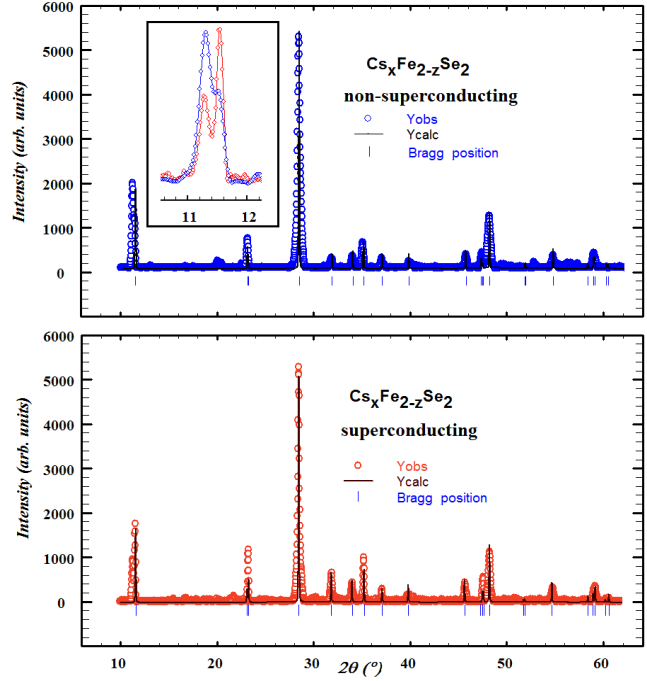

**Figure 3** Powder X-ray pattern of the non-superconducting (blue) and superconducting (red) $Cs_xFe_{2-z}Se_2$ samples. A Cu Kα radiation was used to obtain the X-ray pattern. The inset shows the evolution of the (002) reflection with annealing.

To examine the superconducting properties of $Li_x(C_5H_5N)_yFe_{2-z}Se_2$ and $Cs_xFe_{2-z}Se_2$ the magnetic susceptibility measurements were performed with a Quantum Design PPMS magnetometer on the as-prepared and annealed samples. The temperature dependencies of the real part of the *AC* magnetic susceptibility (measured with a field amplitude $H_{ac}$ = 1 Oe) are presented in Figures 4 and 5, respectively. The *AC* susceptometer was used as insert of the PPMS main unit.

Mössbauer absorbers were prepared in the glove box filled with helium by mixing 25 mg/cm$^2$ of the sample in a powder form with graphite and pressing into pellet. The pellet was covered with a Cytop® polymer to avoid contact with the air. Mössbauer spectra were obtained versus temperature ranging between 4.2 – 300 K for $Li_x(C_5H_5N)_yFe_{2-z}Se_2$ sample in the annealed state. Spectra of the remaining samples were obtained in the temperature range 80 – 300 K. RENON MsAa-3 spectrometer was used with the Kr-filled LND proportional detector. A $^{57}$Co(Rh) commercial source was kept at room temperature. The velocity scale was calibrated by using Michelson-Morley interferometer equipped with the He-Ne laser. Spectra were fitted within transmission integral approximation using GMFPLF application of the Mosgraf-2009 suite [17]. All spectral shifts are reported versus room temperature α-Fe.

## 3. Results and discussion

Figure 4 shows the *AC* magnetic susceptibility data for $Li_x(C_5H_5N)_yFe_{2-z}Se_2$ samples versus temperature obtained at $H_{ac}$ = 1 Oe. The sample in the as-prepared state exhibits some positive and small susceptibility weakly dependent on temperature. This is an indication that some ferromagnetically ordered components are present at all temperatures investigated, and they have very small magnetic anisotropy. A tiny fraction of the sample goes superconducting at about 40 K and one can see barely discernible knee on the susceptibility. Some additional kink is seen at about 8 K due to the traces of the superconducting FeSe [18]. The pattern changes dramatically upon annealing. A positive contribution above superconducting state is almost temperature independent and much smaller. The superconducting transition is clearly visible at the same temperature as previously, but transition is rather broad, as it is not completed until about 25 K. Strong diamagnetism close to the ground state is observable. Figure 5 shows similar results for $Cs_xFe_{2-z}Se_2$ sample. It is interesting to note that beyond superconducting region one has magnetic susceptibility very close to zero and temperature independent. This is an indication of much higher magnetic anisotropy than previously. A transition to the superconducting state occurs for the annealed sample solely at about 25 K. A transition width is somewhat smaller than for pyridine intercalated system.

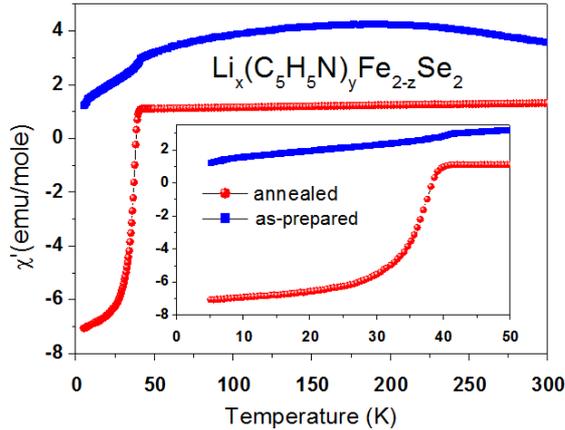

**Figure 4** *AC* magnetic susceptibility of the as-prepared and annealed $Li_x(C_5H_5N)_yFe_{2-z}Se_2$ sample versus temperature with magnetic field amplitude of 1 Oe.

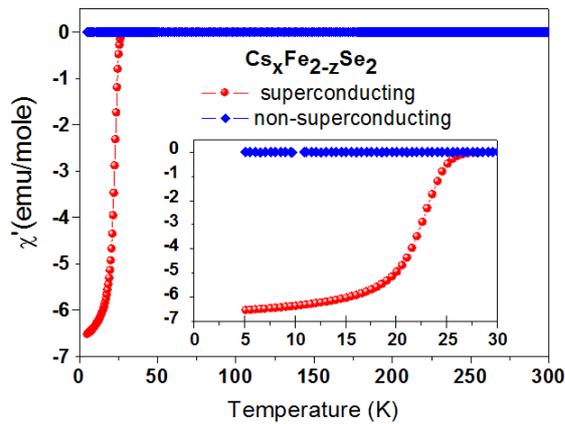

**Figure 5** *AC* magnetic susceptibility of the non-superconducting and superconducting crystals of $Cs_xFe_{2-z}Se_2$ plotted versus temperature with magnetic field amplitude of 1 Oe.

Mössbauer transmission spectra of $Li_x(C_5H_5N)_yFe_{2-z}Se_2$ are shown for several absorber temperatures in Figure 6 for as-prepared and annealed sample. The sample in the as-prepared state shows five spectral components (see Table 1), while annealed sample is characterized by four components as listed in Table 1. The major component is due to iron belonging to the Fe-Se sheets and located in sites being far away from the iron vacancies. This is diamagnetic iron. The second component is likely to be due to isolated trivalent (high spin) iron atoms dispersed between $C_5H_5N$ molecules. The third component has properties of the bulk α-iron and it is likely to originate from the iron precipitates located between Fe-Se sheets. Precipitates have probably very aspherical shape and they are likely to be surrounded by the regular Fe-Se sheets. Hence, they do not contribute to the susceptibility in the superconducting state. The fourth component is likely to be generated by iron atoms located at the surface of above precipitates and having magnetic moment. The last fifth component is due to the iron atoms within Fe-Se sheets, but adjacent to the vacancies, and hence preserving some magnetic moment. Sample annealing changes relative contributions of above components, and the surface iron of precipitates is no longer distinguishable from the bulk iron of the same precipitates. Hence, only four components are left. The hyperfine field on the iron adjacent to vacancies becomes quite large at low temperature excluding these regions from superconducting bulk. Fortunately, such regions are relatively scarce and the superconductivity remains undestroyed.

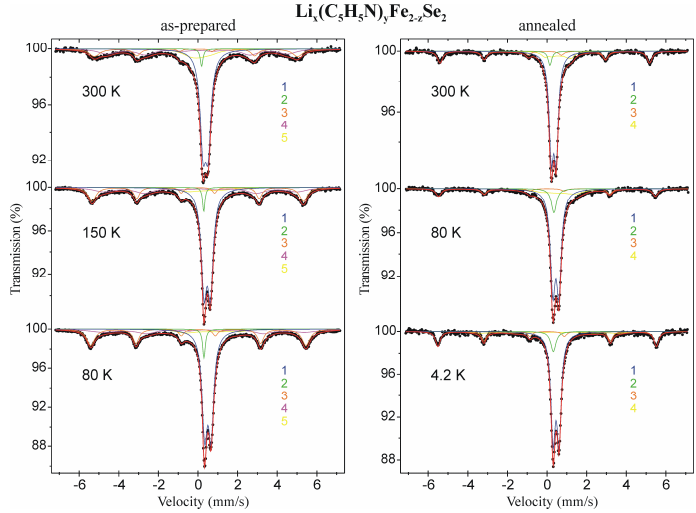

**Figure 6** $^{57}$Fe Mössbauer spectra versus temperature for the as-prepared and annealed $Li_x(C_5H_5N)_yFe_{2-z}Se_2$. Components of the dimensionless cross-sections belonging to various iron sites within the absorber are shown in the same order as in Table 1.

**Table 1**

Mössbauer parameters for as-prepared and annealed $Li_x(C_5H_5N)_yFe_{2-z}Se_2$. The symbol T stands for temperature, $C$ denotes contribution of the particular spectral component, $S$ stands for the total spectral shift versus room temperature α-Fe, Δ denotes quadrupole splitting (including sign for magnetically split spectra), B stands for the hyperfine magnetic field, and Γ denotes absorber line width. Errors for all values are of the order of unity for the last digit shown.

| T (K) | C (%) | S (mm/s) | Δ (mm/s) | B (T) | Γ (mm/s) |
|---|---|---|---|---|---|
| $Li_x(C_5H_5N)_yFe_{2-z}Se_2$ (as-prepared) | | | | | |
| 300 | 52 | 0.492 | 0.25 | - | 0.24 |
|  | 2 | 0.29 | - | - | 0.1 |
|  | 14 | 0.003 | - | 31.83 | 0.14 |
|  | 21 | 0.08 | - | 28.3 | 0.9 |
|  | 11 | 0.15 | 0.69 | 2 | 1.0 |
| 150 | 53 | 0.575 | 0.32 | - | 0.22 |
|  | 3 | 0.40 | - | - | 0.1 |
|  | 19 | 0.101 | - | 33.12 | 0.17 |
|  | 19 | 0.22 | - | 30.3 | 1.0 |
|  | 6 | 0.28 | 0.40 | 6 | 1.0 |
| 80 | 51 | 0.608 | 0.32 | - | 0.20 |
|  | 3 | 0.41 | - | - | 0.1 |
|  | 23 | 0.126 | - | 33.71 | 0.21 |
|  | 16 | 0.44 | -0.05 | 31.65 | 0.6 |
|  | 7 | 0.57 | 0.37 | 8.7 | 1.0 |
| $Li_x(C_5H_5N)_yFe_{2-z}Se_2$ (annealed) | | | | | |
| 300 | 67 | 0.448 | 0.24 | - | 0.16 |
|  | 4 | 0.26 | - | - | 0.12 |
|  | 19 | -0.002 | - | 32.80 | 0.21 |
|  | 10 | 0.71 | -0.07 | 4.2 | 0.33 |
| 80 | 67 | 0.554 | 0.30 | - | 0.18 |
|  | 9 | 0.46 | - | - | 0.21 |
|  | 12 | 0.106 | - | 33.96 | 0.23 |
|  | 12 | 0.84 | -0.26 | 5.5 | 0.97 |
| 4.2 | 66 | 0.565 | 0.30 | - | 0.15 |
|  | 7 | 0.43 | - | - | 0.21 |
|  | 19 | 0.113 | - | 34.21 | 0.18 |
|  | 8 | 0.37 | -0.26 | 21.2 | 0.97 |

Similar situation occurs for $Cs_xFe_{2-z}Se_2$ cesium intercalated material (Figure 7), but the iron environment for respective sites is at least partly different in comparison to the previous material (Table 2). The first and second component in the as-prepared sample is due to iron in the Fe-Se sheets, but the sheets contain either diamagnetic iron (first component) or develop some itinerant magnetic order (second component). Remaining iron seems to be interstitial iron with three discernible environments depending on the degree of clustering. It is definitely in the state close to the state of trivalent high spin iron. Annealing orders Fe-Se sheets leading to the single iron site within a sheet at high temperature. Some of this iron (sheets) develop eventually itinerant magnetism upon lowering temperature. Hence, intercalation by cesium disturbs Fe-Se sheets and one needs further annealing to obtain superconductivity. On the other hand, intercalation by lithium solution in $C_5H_5N$ leads to more ordered structure as even as-prepared material contains some tiny amount of superconductor.

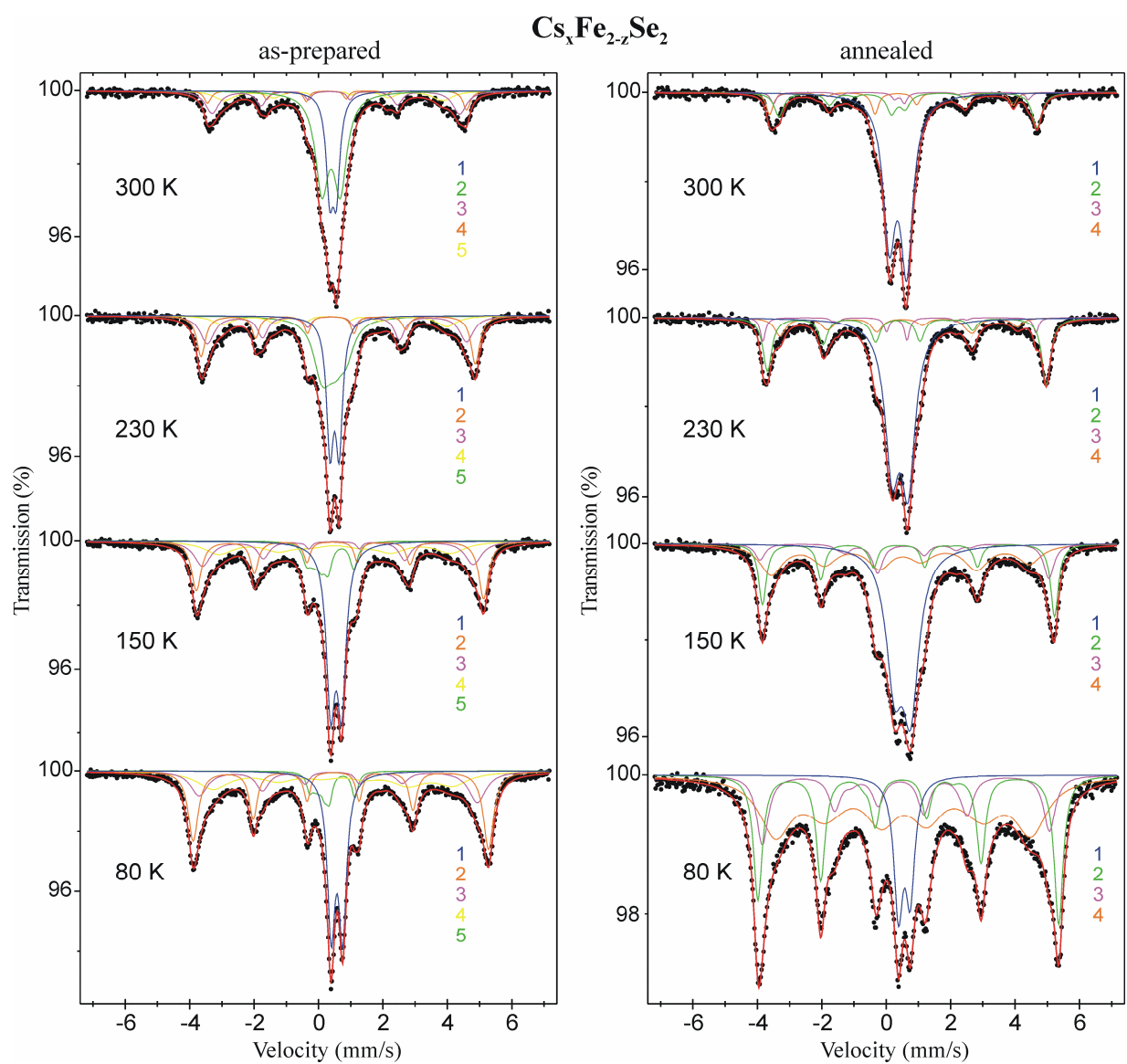

**Figure 7** $^{57}$Fe Mössbauer spectra versus temperature for the as-prepared (non-superconducting) and annealed (superconducting) $Cs_xFe_{2-z}Se_2$. Components of the dimensionless cross-sections belonging to various iron sites within the absorber are shown in the same order as in Table 2.

# Table 2

Mössbauer parameters for as-prepared (non-superconducting) and annealed (superconducting) $Cs_xFe_{2-z}Se_2$. For magnetically split spectra one obtains an angle between the hyperfine field and the principal component of the electric field gradient (assumed to be axially symmetric) as about 45(10)°. The quoted splitting $\Delta$ is calculated as $6A_Q$ with $A_Q$ being the quadrupole coupling constant. Errors for all values are of the order of unity for the last digit shown.

| T (K) | C (%) | S (mm/s) | Δ (mm/s) | B (T) | Γ (mm/s) |
|---|---|---|---|---|---|
| \multicolumn{6}{c}{$Cs_xFe_{2-z}Se_2$ (non-superconducting)} | | | | | |
| 300 | 23 | 0.56 | 0.20 | - | 0.16 |
|  | 42 | 0.50 | 0.57 | - | 0.38 |
|  | 17 | 0.54 | 0.6 | 23.3 | 0.13 |
|  | 7 | 0.52 | 1 | 21 | 0.1 |
|  | 11 | 0.54 | 1 | 20 | 0.2 |
| 150 | 34 | 0.66 | 0.34 | - | 0.22 |
|  | 22 | 0.66 | 0.77 | 26.9 | 0.13 |
|  | 16 | 0.65 | 0.8 | 25.1 | 0.1 |
|  | 18 | 0.61 | 0.9 | 20.9 | 1.0 |
|  | 10 | 0.38 | 0.8 | 3 | 0.19 |
| 80 | 27 | 0.68 | 0.36 | - | 0.18 |
|  | 28 | 0.68 | 0.80 | 27.7 | 0.14 |
|  | 17 | 0.63 | 1.0 | 25.4 | 0.20 |
|  | 21 | 0.71 | 0.6 | 22.6 | 1.1 |
|  | 7 | 0.45 | 0.8 | 3 | 0.1 |
| \multicolumn{6}{c}{$Cs_xFe_{2-z}Se_2$ (superconducting)} | | | | | |
| 300 | 67 | 0.48 | 0.54 | - | 0.35 |
|  | 19 | 0.64 | 1.3 | 19.8 | 0.26 |
|  | 4 | 0.54 | 2 | 16.5 | 0.1 |
|  | 10 | 0.34 | 2 | 12 | 0.2 |
| 150 | 43 | 0.62 | 0.50 | - | 0.50 |
|  | 18 | 0.66 | 0.8 | 27.2 | 0.20 |
|  | 11 | 0.62 | 2 | 20.5 | 0.28 |
|  | 28 | 0.55 | 2 | 15 | 0.8 |
| 80 | 12 | 0.67 | 0.36 | - | 0.24 |
|  | 27 | 0.68 | 1.4 | 27.6 | 0.23 |
|  | 16 | 0.65 | 2.7 | 23.3 | 0.33 |
|  | 45 | 0.65 | 2 | 16 | 1.1 |

Annealing of both materials leads to decrease of the isomer shift for the diamagnetic iron in the Fe-Se sheets. This is an indication of the increased electron density on iron nuclei caused probably by removal/ordering of the iron vacancies within Fe-Se sheets.

Clusters of α-Fe seem to grow upon annealing of the pyridine doped sample at the cost of the iron atoms dispersed between sheets. One observes slight increase in contribution to the spectral area due to the α-Fe as well as increase of the hyperfine field approaching the value of the bulk iron. Some dispersed iron seems to fill iron vacancies in the sheets transforming into diamagnetic form. For cesium intercalated material one observes reduction of the magnetic hyperfine field for iron dispersed between sheets upon annealing. It seems that this fact is due to the increased itinerant electron density between sheets caused by vacancy removal upon annealing.

## 4. Conclusions

Samples despite showing almost single phase X-ray patterns are in fact multi-phase systems on the microscopic scale [19]. There is a superconducting fraction in the regions of the Fe-Se sheets almost free of vacancies, and magnetically ordered fraction in the regions with (ordered) iron vacancies. Some iron occurs between sheets in the form of the high spin trivalent iron. The latter iron tends to form clusters. For large separation of the sheets (the case of lithium/pyridine doping) one sees large enough clusters to form metallic ferromagnetic α-Fe with a significant contribution to the spectrum from the surface iron atoms [20]. It seems that annealing leads to the regions depleted in the high spin intercalated iron and regions enriched in these species. Hence, one has superconducting and magnetic regions within the same material – spatially separated. In order to see hyperfine field on the intercalated isolated iron atoms they have to couple to some ensembles with already ordered magnetic moments e.g. to the magnetically ordered Fe-Se sheets. Nanoscale phase separation was observed in $A_xFe_{2-y}Se_2$ (A=K [21], Rb [22], Tl/K [23]) superconductors by Mössbauer spectroscopy.

In order to obtain superconductivity one needs to order intercalated species, especially for the cesium case. Larger separation of the Fe-Se sheets as provided by pyridine in comparison with cesium leads to increase of the superconducting transition in agreement with the hypothesis of the nearly two-dimensional superfluid density [10].

## Acknowledgments

This work was supported by the National Science Center of Poland, Grant No. DEC-2011/03/B/ST3/00446. A. K.-M. acknowledges the financial support founded by the National Science Center of Poland, Grant No. DEC-2013/09/B/ST5/03391.